\documentclass[vecphys]{svmult}
\usepackage{latexsym}
\usepackage{url}
\usepackage{graphicx}

\usepackage[margin=1in]{geometry}

\mathchardef\mhyphen="2D
\renewcommand{\vec}[1]{\mbox{\boldmath$#1$}}

\newcommand{\bq}{{\vec q}}
\newcommand{\bv}{{\vec v}}

\newcommand{\hmpc}{\,$h^{-1}$\,Mpc}

\newcommand{\LCDM}{$\Lambda$CDM}

\newcommand{\org}{{\scshape origami}}

\newcommand{\citep}{\cite}
\newcommand{\citet}{\cite}

\chardef\til=`\~

\begin{document}

\title*{Tessellating the cosmological dark-matter sheet: origami
  creases in the universe and ways to find them}
\titlerunning{Origami creases in the universe}

\author{Mark C.\ Neyrinck\inst{1}\and
Sergei F.\ Shandarin\inst{2}}
\institute{Department of Physics and Astronomy, The Johns Hopkins University, Baltimore, MD 21218, USA
\texttt{neyrinck@pha.jhu.edu}
\and
Department of Physics and Astronomy, 
University of Kansas, KS 66045, USA
\texttt{sergei@ku.edu}}


\maketitle

\begin{abstract}
Tessellations are valuable both conceptually and for analysis in the
study of the large-scale structure of the universe.  They provide a
conceptual model for the `cosmic web,' and are of great use to analyze
cosmological data.  Here we describe tessellations in another set of
coordinates, of the initially flat sheet of dark matter that gravity
folds up in rough analogy to origami.  The folds that develop are
called caustics, and they tessellate space into stream regions.
Tessellations of the dark-matter sheet are also useful in simulation
analysis, for instance for density measurement, and to identify
structures where streams overlap.
\end{abstract}


\section{Introduction: the cosmological dark-matter sheet}
In Einstein's theory of general relativity, gravity comes about
through the distortion that matter and energy produce in the
four-dimensional manifold of spacetime.  Gravity also causes another
manifold that pervades spacetime to distort and fold: the sheet of
dark matter.

Just after the big bang, the matter was almost uniformly distributed,
i.e.\ the density varied very little from point to point in space.
These tiny density fluctuations are thought to be random quantum
fluctuations that were `inflated' in the first instants to macroscopic
size.

It is useful to think of the matter occupying vertices of a regular
mesh, and to represent the density fluctuations as small distortions
of this mesh.  Where there is a bit more matter than average, the mesh
has contracted, and where there is less matter than average, the mesh
has expanded.

As particles move around in three-dimensional space, they occupy
worldlines in four-dimensional spacetime.  It is also useful to think
of particle trajectories in yet another, six-dimensional `phase space'
of position and velocity.  Each particle in the universe can be
plotted in this 6D phase space, three of the dimensions given by its
spatial position, and three by its velocity.

In phase space, the primordial state of the universe was well
characterized by particles separated from each other in position
dimensions, but with little separation in velocity (all velocities,
after subtracting out the universe's expansion, were nearly zero).  In
this sense, the matter sheet was `flat.'  The coordinates of a
particle within the initial, flat sheet is called its {\it Lagrangian}
position; this remains constant for a given particle for all time.
The usual position of a particle as it moves around in space is called
its {\it Eulerian} position.

As time passes, the largest effect on the matter sheet is stretching
from the expansion of the universe, which generally increases the
physical spatial separation of particles.  As is usually done in
cosmology, however, we use comoving coordinates, i.e.\ we divide out
the expansion.  The comoving coordinates of particles at rest with
respect to the expansion of the universe do not change.

Gravity also amplifies the small distortions in the matter-sheet
sheet, increasing velocities.  In fact, the dark-matter sheet has
special physical significance in general relativity: it is the set of
observers (dark-matter particles) that have been freely falling in
gravity since the big bang.  This follows from the (astronomical)
definition of dark matter as matter that interacts only through
gravity.

The growth of structure in the universe is a balance between gravity
pulling matter together and the expansion of the universe damping such
motions, as seen in comoving coordinates.  Nevertheless, in
`overdense' regions where the sheet has contracted, more matter
accumulates, so the sheet contracts further.  Likewise, underdense
regions repel matter, expanding the sheet to form a void.

In overdense regions, the sheet eventually bunches together and folds.
Gas (normal matter) collects in these regions, and forms galaxies.
When two gas streams encounter each other, they collide and shock, to
a good approximation forcing a unique gas velocity at each point.

The dominant form of matter in the universe, dark matter, on the other
hand, only interacts gravitationally.  Two dark-matter particles
encountering each other at the same position do not collide, i.e.\ do
not change their trajectories.  (In many models, dark matter can very
rarely collide, but here we neglect this possibility.)  Often a single
point of Eulerian space can have dark matter flowing with many
discrete velocities.  Thus (given the spatial continuity of the
mapping from initial to final positions), the dark-matter sheet has
folded up, if considered in 6D phase space.  Since particles cannot
have the same positions and velocities without also being initially
coincident, the mesh cannot intersect itself.

\begin{figure}
  \begin{center}
    \includegraphics[scale=0.7]{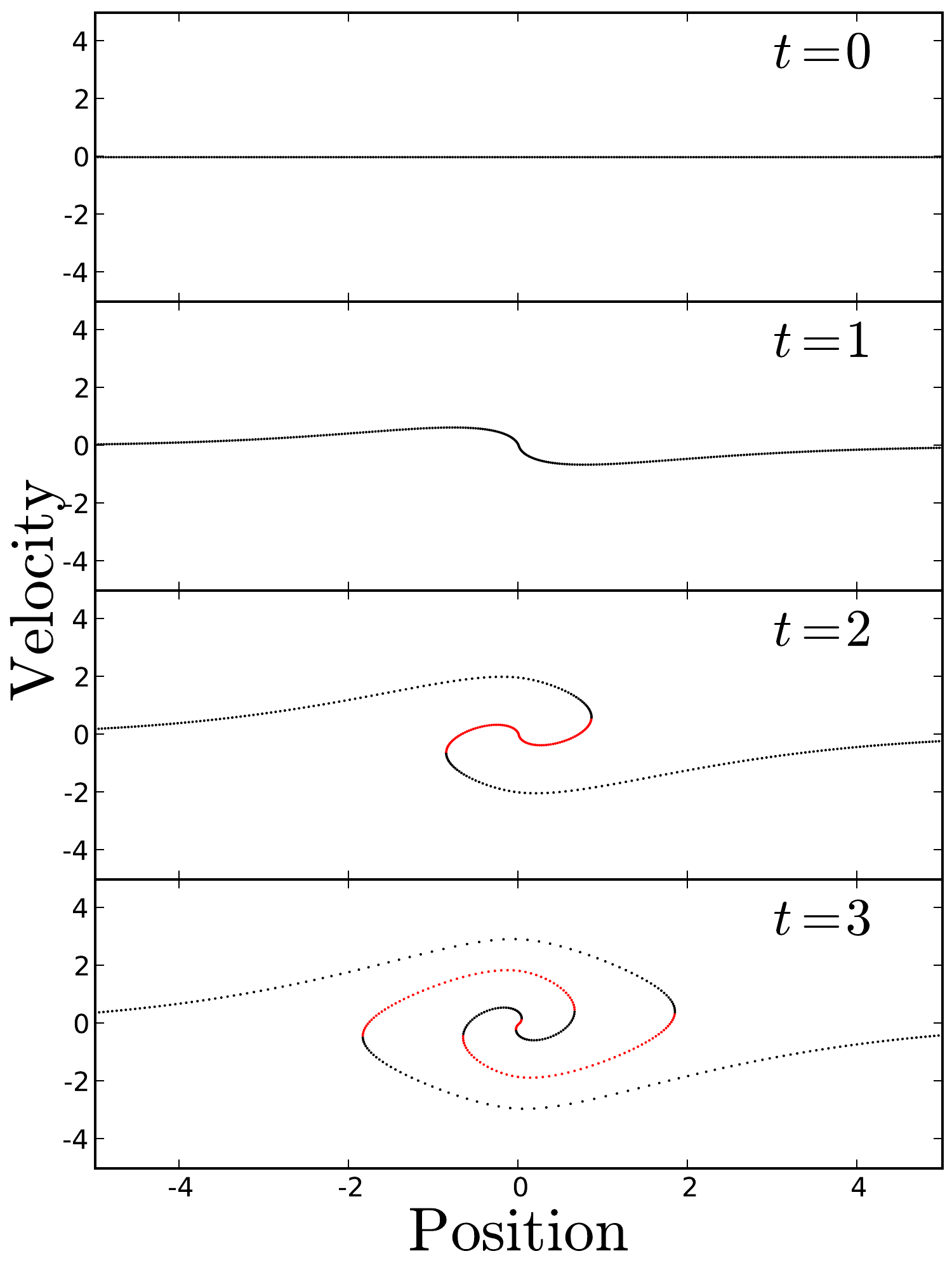}
  \end{center}  
  \caption{A schematic phase-space spiral that corresponds to the
    collapse of a `halo' in a one-dimensional universe.  Patches
    oriented forwards in position space (i.e. projecting down to the
    $x$-axis) are colored black, while patches oriented backwards are
    red.  These two possibilities make the set of streams (contiguous
    patches with the same orientation) two-colorable.  All
    one-dimensional sets of contiguous regions are two-colorable, but
    when we go to two and three dimensions, this property becomes
    quite special.
    \label{fig:spirals}
  }
\end{figure}

A schematic example of halo collapse in a one-dimensional universe
(with therefore a two-dimensional position-velocity phase space) is
shown in Fig.\ \ref{fig:spirals}.  Particles start out equally
separated, but are drawn into the center, their Lagrangian string
winding up into a spiral.  The quasi-circularity of the spiral comes
from particles oscillating back and forth about the center of the
potential.  Different orientations of the initially flat string of
dark-matter particles are colored black (forward) and red (backward).
Contiguous regions on the string that are oriented the same way are
called {\it streams}.  The boundaries between streams, folds in the
string when projected down to the $x$-axis, are called {\it caustics}.
At caustics, in the limit of infinitesimal particles and infinite
spatial resolution, the densities become infinite.  Thus they may
greatly enhance the chances of observing dark matter, which may
collide observably in very dense environments
\citep{Hogan2001,NatSik2008,VogelsbergerWhite2011}.

The folding in this figure suggests an analogy to origami: indeed,
cosmic structure is built out of origami-like folds, although there
are some differences in detail.  The rest of this contribution is
organized as follows.  First, in Sec.\ \ref{sec:origami}, we describe
some mathematical propeties of paper origami, and how to understand
origami crease patterns in terms of tessellations.  In
Sec.\ \ref{sec:oricosmi}, we explain the analogy between paper and
cosmological origami in detail.  Then, in
Sec.\ \ref{sec:euleriancaustics}, we give some properties of the
structures gravity builds from origami-folding up the dark-matter
sheet.  In Sec.\ \ref{sec:lalgorithms}, we describe how various
tessellations of the dark-matter sheet in Lagrangian space can be used
to analyze and extract useful information from cosmological
simulations.

\section{Origami mathematics}
\label{sec:origami}

\begin{figure}
  \begin{center}
    \includegraphics[scale=0.43]{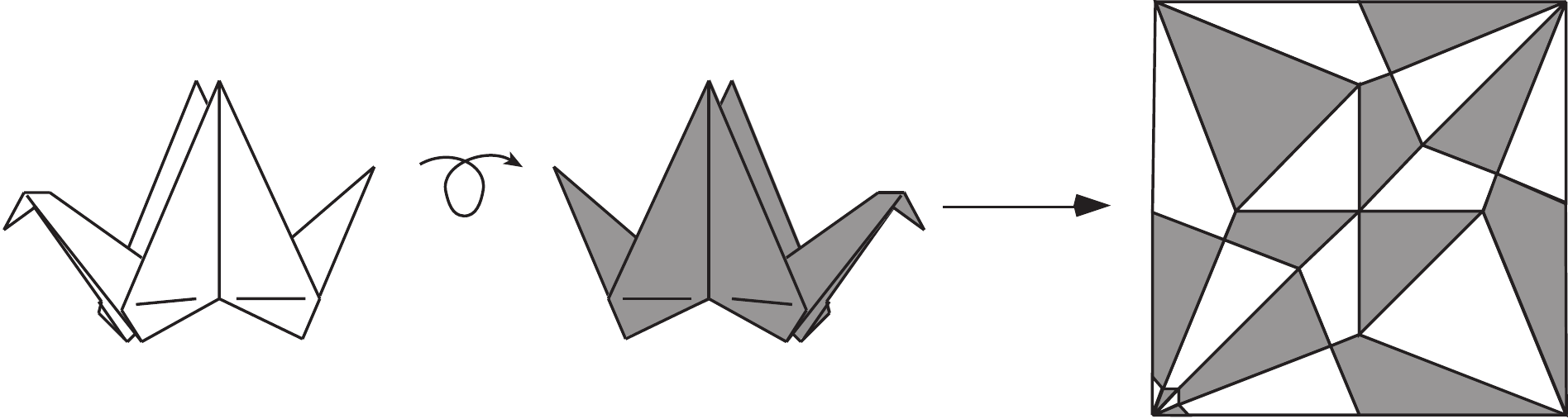}
  \end{center}  
  \caption{Two-coloring of the polygons outlined by creases in an
    origami `traditional Japanese flapping bird' (similar to a
    crane). Polygons facing `up' (out of the page in the leftmost
    diagram, with the head facing to the left) are colored white,
    while polygons facing `down' (into the page in the leftmost
    diagram) are painted gray.  The crane has been unfolded for the
    rightmost diagram.  Creases are shown as lines here; the polygons
    outlined by them may be colored with only two colors, such that
    polygons straddling creases never have the same color.  Figure
    from \citet{Hull2006}, courtesy Tom Hull.
    \label{fig:twocolor}
  }
\end{figure}

There has been much mathematical work in origami, most of it recent in
the ancient history of the art form
\cite{Row1966,Martin1998,Lang2003,Hull2006}. `Flat origami' is the
class of origami most easily relatable to large-scale structure.  In
flat origami, folding of a two-dimensional sheet is allowed in three
dimensions, but the result is restricted to lie flat in a plane,
i.e.\ it could be squashed between pages in a book without acquiring
any new creases.  The class of flat-foldable origami is quite large,
for example encompassing the paper crane, similar to the model shown
in Fig.\ \ref{fig:twocolor}.

There are several theorems that have been proven about flat origami
\citep{Hull1994,Hull2002}.  One is the two-colorability of polygons
outlined by origami crease lines, as shown in Fig.\ \ref{fig:spirals}
in one dimension.  Two colors suffice to color them so that no
adjacent polygons share the same color.

To see why two colors suffice, consider the bird in
Fig.\ \ref{fig:twocolor}.  Both sides of it are shown, along with its
appearance when unfolded.  Each polygon is colored white or gray
according to whether the polygon is facing `up,' i.e.\ with the same
orientation as it did initially, or `down,' if it has been flipped
over.  This uniquely colors each polygon, and each crease does indeed
divide `up' from `down' polygons.  According to the four-color theorem
(e.g.\ \cite{wilson2002graphs}), a general set of planar regions is
colorable by four colors. So, the ability to produce a flat-foldable
origami design from a crease pattern reduces the so-called {\it
  chromatic number} (the number of colors necessary such that
neighboring regions are not colored the same) from four to two.

A work of flat origami can be thought of as a function (specifically,
a continuous piecewise isometry) mapping the unit square (the unfolded
sheet at right in Fig.\ \ref{fig:twocolor}) into the plane.  Each
crease produces a reflection, reversing the direction of the vector on
the paper perpendicular to the crease.  The function is defined on
each polygon by a sequence of these reflections.  The color in each
polygon corresponds to its parity, i.e. depending on whether the
number of reflections used to define the function on that polygon is
odd or even.  The parity can also be measured locally with the
determinant of the matrix defining the function on the polygon; we
will also use this latter definition in the cosmological case below.

Besides two-colorability, there are other properties that
flat-foldable crease patterns have.  For example, Maekawa's theorem
states that in a flat-foldable crease pattern, the numbers of
`mountain' and `valley' creases around a vertex differ by two.  (A
mountain crease becomes folded to form an upward-pointing ridge; a
valley crease is folded in the opposite way.)  Maekawa's theorem is
likely to be applicable to a stretchable 3D cosmological sheet, as
well, in a more complicated form, but we have not investigated this
possibility. 

Even for paper origami, a difficult problem is to test
that an arbitrary crease-pattern is physically flat-foldable without
the paper intersecting any folds; this is an NP-complete problem
\citep{BernHayes1996}.  There are further results that, for instance,
describe the angles around vertices, but they depend on the
non-stretchability of the origami sheet, making them inapplicable to
the cosmological case.

\section{Origami large-scale structure}
\label{sec:oricosmi}
Moving from paper origami to cosmological structure formation
introduces a few changes.  The manifold (sheet) has three instead of
two dimensions.  It folds in six dimensions (three position, and three
velocity) instead of three.  The sheet also stretches inhomogeneously,
stretching more in voids than in dense regions.  In dense regions, it
can also stretch violently in the velocity dimensions.

\begin{figure}
  \begin{center}
    \includegraphics[scale=0.3]{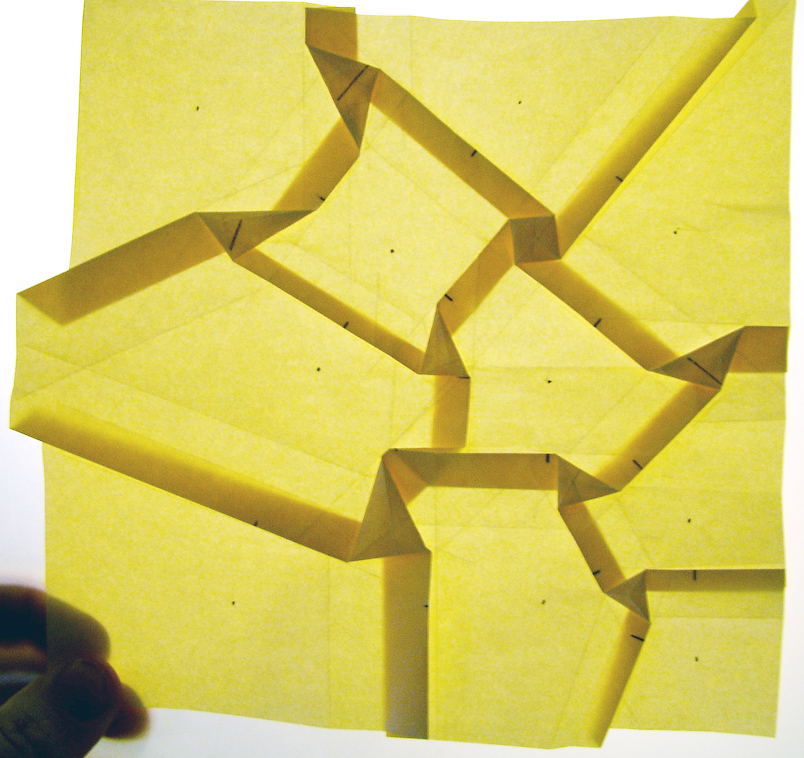}
  \end{center}  
  \caption{A Voronoi origami tessellation that resembles cosmological
    voids, filaments and haloes in a two-dimensional universe.  If the
    work were unfolded, the polygons outlined by creases would be
    two-colorable according to which way the polygon is facing.
    E.g.\ `voids' and the topmost polygons in `haloes' could be
    colored black, and the paper turned upside down within the
    `filaments' could be colored white.  Design and photo 
    by Eric Gjerde ({\tt http://www.origamitessellations.com/}), used
    with permission.
    \label{fig:vorigami}
  }
\end{figure}

In spite of these differences, it is still possible to approximate two-dimensional large-scale structures with origami designs.
Fig.\ \ref{fig:vorigami} shows a work of flat origami that bears some
resemblance to the cosmic web of filaments and clusters in cosmology.
The `voids' are Voronoi cells generated from black pencil-marks on the
paper.  Voronoi models of large-scale structure are good heuristic
models of cosmological structure formation
\cite{IckevandeWeygaert1987,KofmanEtal1990, HiddingEtal2012}.  The
present figure corresponds most closely to a Zel'dovich-approximation
evolution of particle displacements, in which structures fold up when
expanding voids collide, but overshoot and do not undergo realistic
further collapse.

\begin{figure}
  \begin{center}
    \includegraphics[scale=0.45]{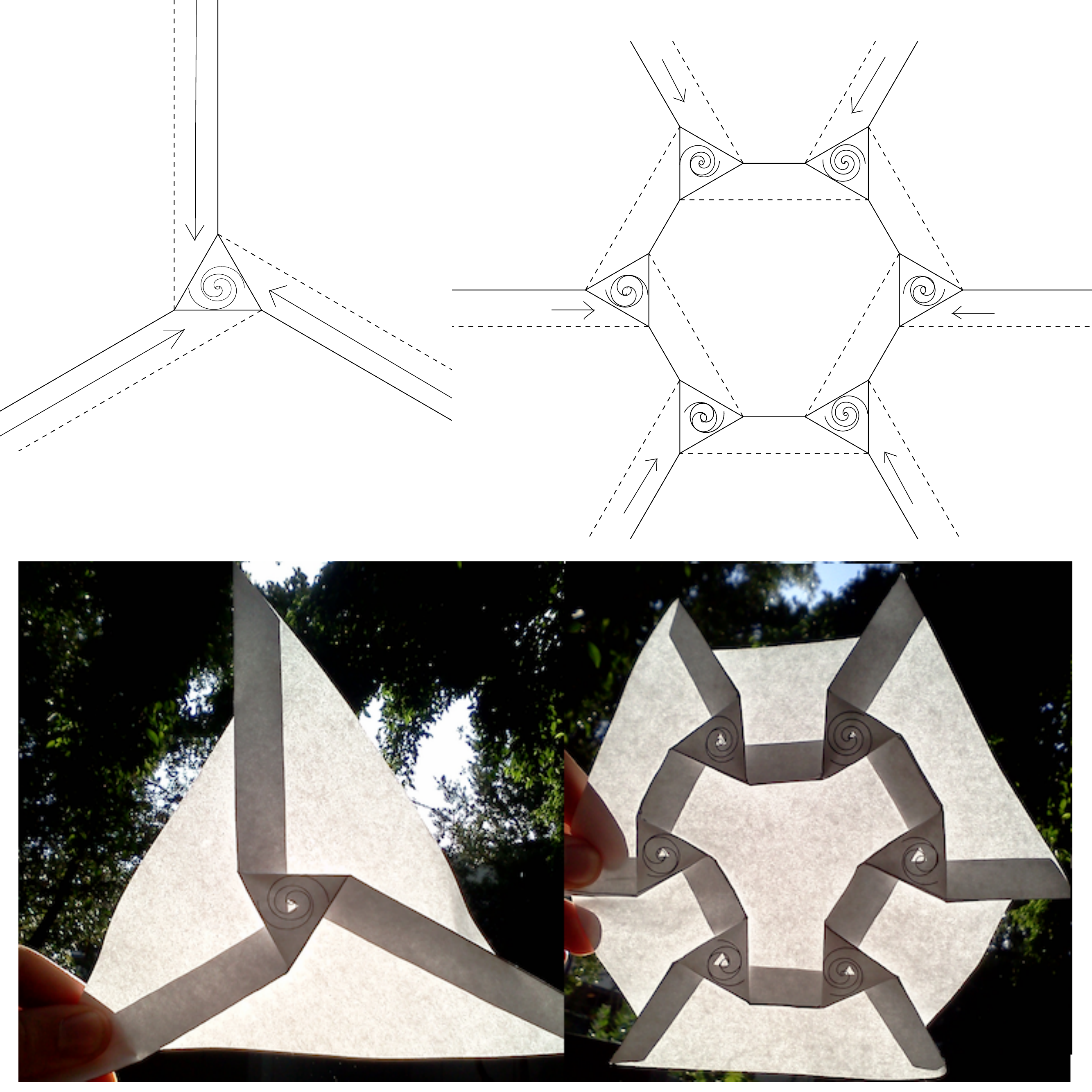}
  \end{center}  
  \caption{Top, crease patterns that fold up into schematic `galaxies'
    surrounded by filaments, at bottom.  In the crease patterns, solid
    lines give mountain folds, folding to form a ridge, and dashed
    lines give valley folds.  The arrows indicate the direction of
    matter flow within the folded-up filaments.
    \label{fig:origalaxies}
  }
\end{figure}

Fig.\ \ref{fig:origalaxies} shows two crease patterns: one that folds up
into a single schematic `galaxy,' and one that folds up into a
hexagonal void surrounded by six galaxies.  They have proven useful at
public-outreach events, and are available at
\url{http://skysrv.pha.jhu.edu/~neyrinck/origalaxies.html}.  These
designs are based on elements of Eric Gjerde's `Tiled Hexagons'
pattern, in his {\it Origami Tessellations} book \cite{Gjerde2008},
which contains several other interesting origami tessellations.

These figures are interesting pedagogically, but also suggest a reason
why filaments are so common in the universe.  Here, galaxies, or
knots, {\it cannot} form without associated filaments.  This
requirement comes from the non-stretchability of origami paper (a
property which the dark-matter sheet lacks), but it still suggests a
strong tendency for filaments to form along with galaxies.  It also
suggests a reason why galaxies tend to accrete much angular momentum:
it seems to be easier to fold a galaxy when the filament has a nonzero
impact parameter with respect to the center of the galaxy.  When the
accreting matter arrives in the galaxy, it then torques it up.  These
are not proofs, but interesting suggestions.

Moving toward more mathematical rigor, we now consider local parities
of patches on the cosmological sheet. The parity may still be defined
in the same way as in flat origami, and it may have only one of two
values (positive or negative).  The parity at a particle is measurable
from how the particles initially adjacent to it have distorted around
it.  Mathematically, the parity is the sign of the determinant of the
deformation tensor that takes initial to final coordinates; see for
example \cite{whi09,Neyrinck2012b} for details.

\subsection{Streams and caustics in Lagrangian space}
The dark-matter sheet folds up in Eulerian space in an overwhelmingly
rich way, visually corresponding somewhat to rococo art.  Beautiful
figures of the structure that develops can be seen in
\citet{VogelsbergerWhite2011}, for example.

We defined caustics and streams for a 1D universe around
Fig.\ \ref{fig:spirals}, and use the same definitions in a 3D
universe.  A stream is a contiguous three-dimensional region with the
same orientation, or parity.  A caustic is a two-dimensional surface
separating streams from each other.  Defined this way, a caustic
indeed corresponds to a fold, since the parity swaps if one moves
across it.

By definition, then, Lagrangian space (the dark-matter sheet) is
tessellated by streams that are outlined by caustics.  In addition,
the streams are two-colorable, just as in paper origami.  This is
because the parity may take only two values, and changes as one
crosses a caustic along the sheet.

Two-colorability may seem hopelessly academic, and indeed it does not
have obvious observational consequences.  But in fact it greatly
restricts the arrangement of streams on the unfolded dark-matter
sheet.  A tessellation in greater than two dimensions has in principle
no bound on its chromatic number (the number of colors required).  In
relation to the famous four-color theorem (four colors suffice to
color any planar arrangement of regions), Guthrie \cite{Guthrie1880}
discussed the impossibility of restricting the chromatic number of an
arrangement of solid regions in three dimensions.  He constructed a
set of arbitrarily many long sticks, each of which touches all others.
Such an arrangement is possible, for example, if each stick is
slightly rotated from its neighbor.  Flexible sticks are especially
capable of touching all the others; imagine a bowl of arbitrarily long
spaghetti noodles.  In this case, the chromatic number is bounded only
by the number of sticks.

In graph theory, a two-colorable graph is called {\it bipartite}.
Using the graph-theory terms of `vertices' that are linked by `edges,'
the vertices of the cosmological bipartite graph are the
three-dimensional stream regions, and the edges are the caustic
surfaces between them.

At least one whole book is devoted to the subject of bipartite graphs
\citep{BGraphs1998}; here we list a few of their properties,
translating into cosmological terms.  First, there is no path
(stepping from stream to stream through caustics) starting and ending
at the same stream that consists of an odd number of steps.  Another
result, K\"{o}nig's Minimax Theorem, pertains to the necessity of
caustics to form streams.  It states that the minimum number of
streams needed to touch all caustics with streams equals the maximum
possible number of caustics involved in a matching between streams of
opposite parity.  A `matching' is a set of caustics linking pairs of
streams, such that no stream is touched by more than one caustic.
Considering the dual graph, in which streams and caustics swap roles,
K\"{o}nig has another result.  His Coloring Theorem for bipartite
graphs applies to the dual graph of caustics joined by streams: the
chromatic number for the dual graph equals the maximum number of
caustics around a single stream.  If the graph of streams joined by
caustics were not bipartite, the dual graph would generally have a
larger chromatic number than the maximum number of caustics around a
single stream.

We close this section with a technical caveat about this
stream-caustic definition.  In principle, the stretchable cosmological
sheet can fold in a way that is impossible for paper origami.  Unlike
paper folds, cosmological caustics can form through spherical or
cylindrical collapse, not just planar collapse.  Spherical collapse
reverses parity just as in planar collapse, but cylindrical collapse
does not; it simply produces a 180$^\circ$ rotation in the two axes
perpendicular to the cylinder.  However, in a physically realistic
situation, the probability that more than one axis will collapse
exactly simultaneously is zero, so we adopt the view that caustics
always form one-at-a-time, if viewed at sufficiently high resolution.

\subsection{Simulation measurements}
\begin{figure}
  \begin{center}
    \includegraphics[scale=0.4]{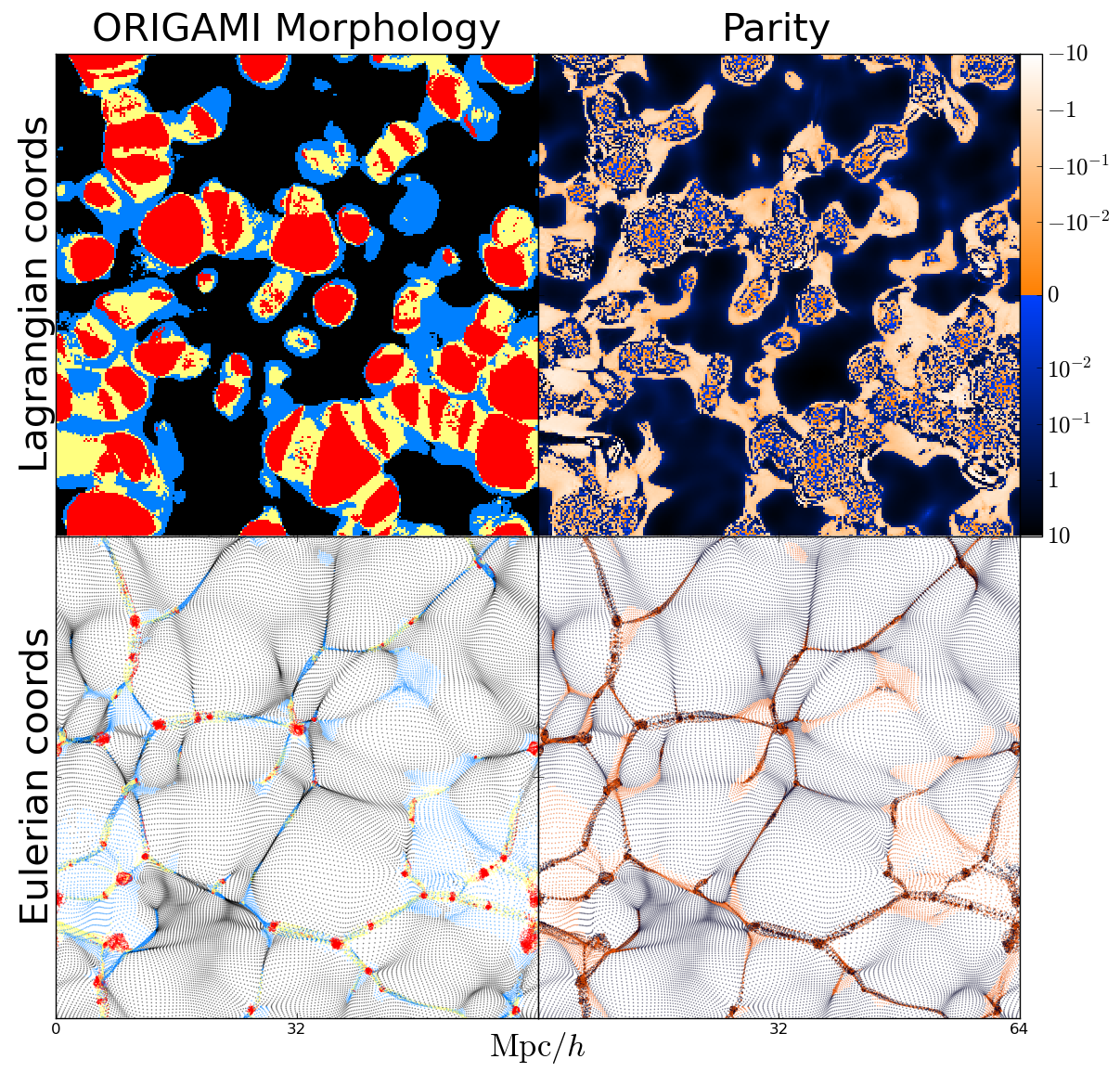}
  \end{center}  
  \caption{The folding of a cosmological sheet (top, unfolded;
    bottom, folded). Quantities were measured from a 2D sheet
    of particles from a 3D \LCDM\ $N$-body simulation.
    The $256^2$ particles share the same $z$-coordinate in the
    initial-conditions lattice, where $z$ points out of the
    page. Before running the simulation, the initial conditions were
    smoothed with a 1\hmpc\ Gaussian window, to inhibit small-scale
    structure formation.  Top panels use Lagrangian coordinates, in
    which each particle is a square pixel in a 256$^2$-pixel image.
    In the bottom panels, particles are shown in their actual
    present-epoch Eulerian $(x,y)$ coordinates, projecting out the
    $z$ coordinate (in which the slice does have some extent).  In
    left-hand panels, void, wall, filament and halo
    \org\ morphologies are shown in black, blue, yellow and red,
    respectively.  In right-hand panels, particles are colored
    according to $J$, i.e.\ the volume of their fluid element times
    its parity.  Black/blue particles have right-handed parity (as
    in the initial conditions), and white/orange particles have
    swapped, left-handed parity.}
  \label{fig:morphparity1}
\end{figure}

Fig.\ \ref{fig:morphparity1} shows the folding up of a 2D cosmological
sheet of particles from a \LCDM\ (the current, observationally
successful model of cosmology) 3D gravitational simulation, with its
initial small-scale fluctuations dampened for clarity. In the top
panels, pixels represent particles in the square grid of Lagrangian
coordinates; in origami terms, this is the flat sheet before
folding. The bottom panels show particles in Eulerian coordinates.

The `morphology' of the left panels describes whether they are void,
filament, sheet, or halo particles.  This morphology is measured using
the \org\ \cite{FalckEtal2012} algorithm discussed above (not to be
confused with the origami analogy itself).

In the right panels, particles are colored primarily by parity
(white/orange or black/blue).  Particles which have been flipped by
caustics an even number of times (including zero) and have the
original, right-handed orientation are black/blue; particles that have
been flipped an odd number of times and have left-handed orientation
are white/orange.  In the finer color gradation (along the
white/orange or black/blue spectra), the upper-right panel
additionally shows the magnitude of the volume each particle occupies
on the dark-matter sheet (inversely proportional to its density).
Note that the magnitude is quite small in the cores of halo regions,
because mass elements shrink considerably in high-density halo
regions.

There is some agreement between outer caustics identified by
\org\ morphology (the boundaries between black and non-black regions)
and as measured by parity (the outermost boundaries between dark blue
and light orange regions), but the agreement is not perfect.  See
\cite{Neyrinck2012b} for further discussion and details.

\begin{figure}
    \begin{center}
      \includegraphics[scale=0.4]{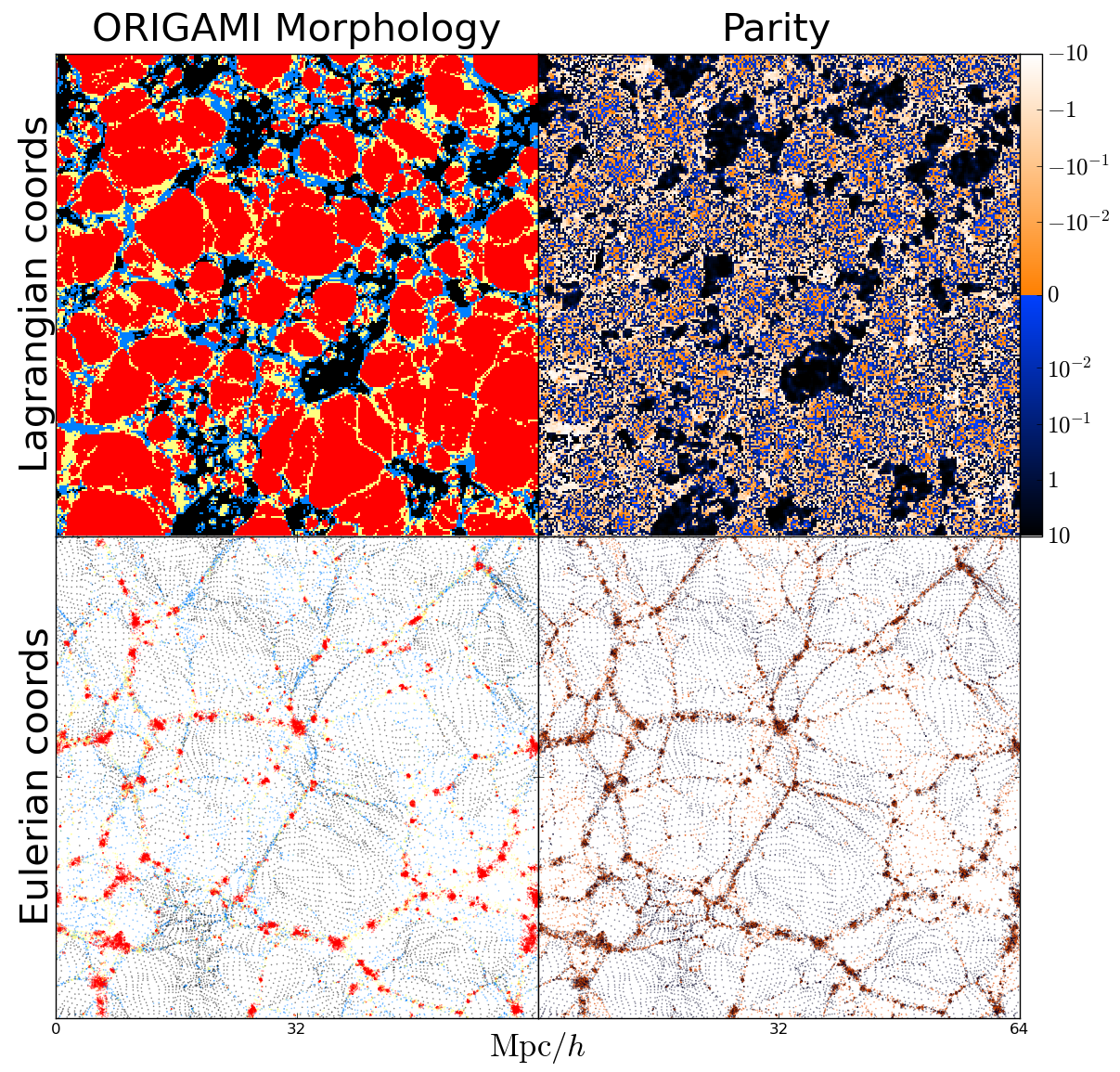}
    \end{center}  
    \caption[1]{Same as Fig.\ \ref{fig:morphparity1}, except measured
      from a simulation with full initial power, i.e. unsmoothed
      initial conditions.}
    \label{fig:morphparity0}
\end{figure}

Fig.\ \ref{fig:morphparity0} is the same figure as
Fig.\ \ref{fig:morphparity1}, except that the simulation was run
without smoothing the initial conditions, causing the true amount of
small-scale structure to appear.  This structure in
Fig.\ \ref{fig:morphparity0} makes the parity map (the upper-right
panel) much more cluttered.  There are many visible extended streams
(patches of identical parity), especially those that correspond to
large voids, but much of the plot looks essentially random.  Note also
here that the Lagrangian patches that correspond to voids are much
smaller than in the previous figure, indicating that the dark-matter
sheet has stretched more.  Given the near-randomness of the parity
deep within haloes, it seems that particles here have crossed many,
many caustics.  We caution, however, that some of this apparent
randomness could be `noise' from finite resolution.

\section{Properties of structures built from folds in the dark-matter sheet}
\label{sec:euleriancaustics}
\begin{figure}
  \begin{center}
    \includegraphics[scale=0.3]{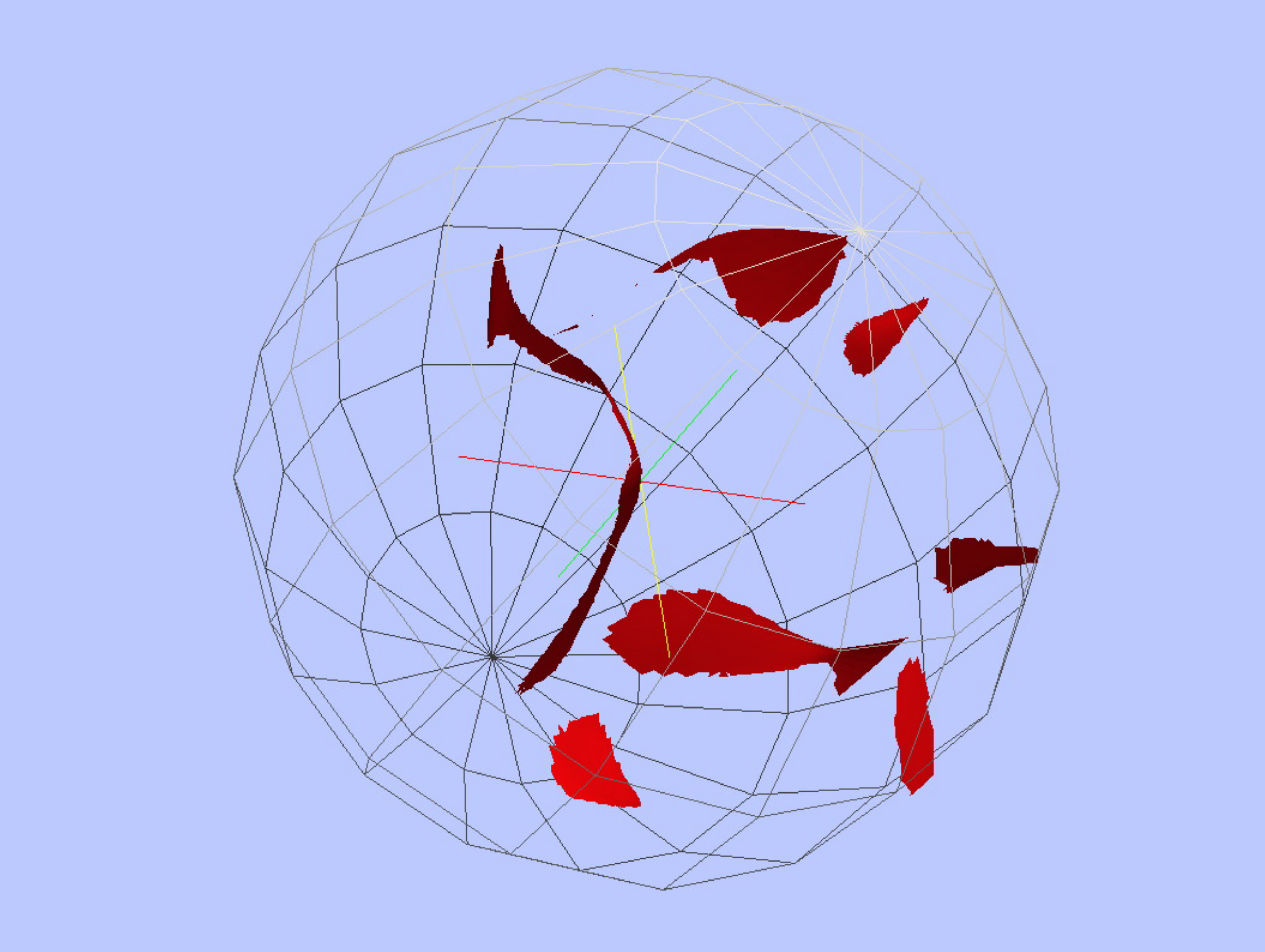}
  \end{center}  
  \caption{A few isolated pancakes bounded by caustics are shown at the early stage of
        the evolution described by Eq.\ (\ref{eq:za}).
    \label{fig:pancakes}
  }
\end{figure}

\begin{figure}
  \begin{center}
    \includegraphics[scale=0.3]{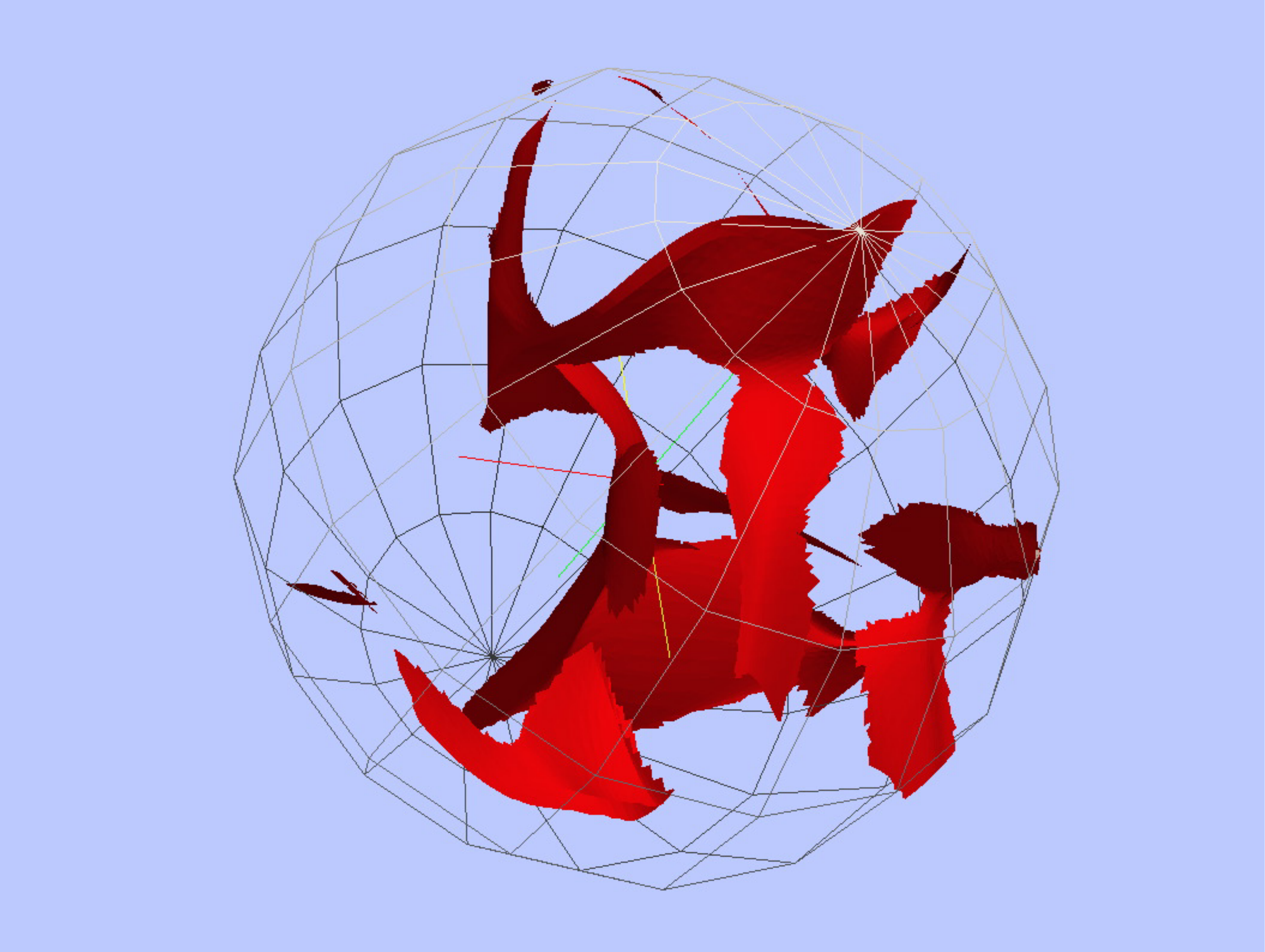}
  \end{center}  
  \caption{Pancakes have grown in size and some of them merged into a large connected structure.
    \label{fig:structure}
  }
\end{figure}

Now we turn our attention to the structures built up by caustics in
the dark-matter sheet. Zel'dovich \cite{Zeldovich1970} predicted the
formation of caustics at the stage when the evolution of the density
field reaches non-linearity. Fig.\ \ref{fig:pancakes} illustrates the
beginning of the structure formation when a few very thin
concentrations of mass emerged. It is worth noting that in order to
avoid blocking the view only a relatively small sphere cut from the
large volume is shown.  The surfaces seen in the figure are the
caustic surfaces where density is formally infinite.  A very thin
layer of highly compressed matter between two caustics form the first
nonlinear structures which Zel'dovich called pancakes.  Initially each
pancake consists of three streams of mass moving with different
velocities through each other, as illustrated in
Fig.\ \ref{fig:pancakes}.  As time passes, pancakes grow in size,
merge with other pancakes and develop an intricate structure shown in
Fig.\ \ref{fig:structure}. The number of streams in pancakes rapidly
grows, and in addition, filaments and compact halos emerge.  If the
flow has no curl -- a condition which holds to a good approximation
except in highly nonlinear regions -- Arnold rigorously proved that
only six generic types of singularities exist.  According to Arnold's
ADE classification, these are $A_2$ (surfaces shown in Figures
\ref{fig:pancakes} and \ref{fig:structure}) and $A_3$ (lines seen as
the contours of pancakes in those figures).  The remaining four types
occur in isolated points: $A_4$ and $D_4$, that persist for some
finite length of time, and $A_5$ and $D_5$, that exist only
instantaneously.  There are also subclasses in some of these classes
but the details are not important for the current discussion.

In the Zel'dovich approximation \cite{Zeldovich1970, ArnoldEtal1982,
  ShandarinZeldovich1989} the singularities can be found directly
from the initial velocity field $\bv(\bq) = - \nabla \Phi(\bq)$ that
completely determines the evolution via a simple map 
\begin{equation}
  x_i(q,t) = q_i + D v_i,~~~ v_i \equiv {d\, x_i \over d\,D} = v_i(q).
  \label{eq:za}
\end{equation}
Here we use so-called comoving coordinates, that exclude the
uniform expansion of the universe.  The function $D = D(t)$
monotonically increases with physical time, and thus can be used as a
time coordinate.  The coordinates $x_i$ and $q_i$ are the Eulerian and
Lagrangian coordinates of the fluid particles.  The volume of a fluid
element can be found from the continuity equation and can be
conveniently expressed in terms of the eigenvalues $\lambda_1(q),\,
\lambda_2(q), \, \lambda_3(q)$ of the deformation tensor $d_{ik}
\equiv - \partial v_i / \partial q_k$:
\begin{equation}
      V(q,t) = V_0 (1-D\lambda_1)(1-D\lambda_2)(1-D\lambda_3).
      \label{eq:volume}
\end{equation} 
The volume collapses to zero when one factor in parentheses in the
above equation vanishes. At this instant of time the fluid particle is
squashed into one of two dimensional surface elements comprising the
caustic surface and then expands with different parity.  If one
assumes that three eigenvalues at each point are ordered
[$\lambda_1(q) \ge \lambda_2(q)$ and $ \lambda_2(q) \ge \lambda_3(q)$]
then the first pancakes arise around maxima of $\lambda_1(q)$.

\begin{figure}
  \begin{center}
    \includegraphics[scale=0.3]{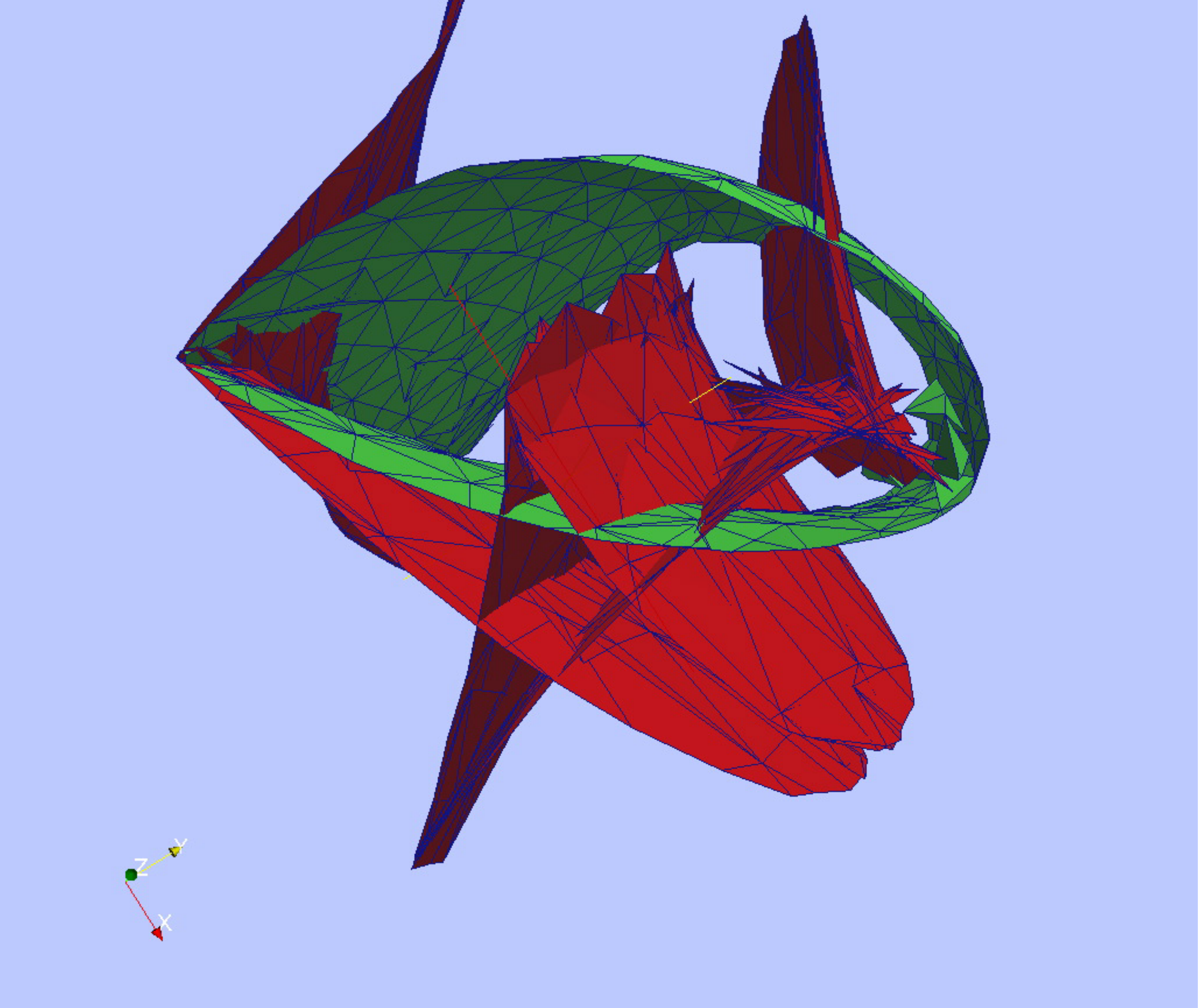}
  \end{center}  
  \caption{Caustics related to two eigenvalues are shown by different colors. The nice smooth
        edges of caustics are due to cutting by an enclosing sphere (not shown).
    \label{fig:eig12}
  }
\end{figure}

For a while only the caustics of $A$ types, related to $\lambda_1$,
occur. Then $A$-singularities related to $\lambda_2$ and $\lambda_3$ as
well as $D$-singularities related to points where
$\lambda_1(q)=\lambda_2(q)$ or $\lambda_2(q)=\lambda_3(q)$ arise.  It
is worth remembering that the points where all three eigenvalues would
have the same value do not exist in a generic field.  Figure
\ref{fig:eig12} shows two types of caustics (red and green) related to
two eigenvalues in a small sphere (not shown).

Even a relatively simple evolution described by the Zel'dovich
approximation results in a very complex structure of caustics
characterized by numerous intricate crossings. Unfortunately the
Zel'dovich approximation can be used for a qualitative or crude
quantitative analysis of the structure in the universe only at early
nonlinear stages, although straightforward modifications can improve
some aspects of the approximation
\cite{ColesEtal1993,SahniColes1995,Neyrinck2012b}. The more realistic picture
emerging from cosmological N-body simulations shows that the
complexity of the structure grows with increasing rate in the
concentrations of mass, in particular in halos
\cite{VogelsbergerWhite2011}.

The presentation of the N-body results in the form of particle plots
is a far more popular in cosmological literature than other types of
illustrations.  Unfortunately the pictures of the particle
distributions fail to reveal caustics in N-body simulations except
for the most massive caustics in the simulations of a single halo with
billions of particles \cite{VogelsbergerWhite2011}.

\begin{figure}
  \begin{center}
    \includegraphics[scale=0.3]{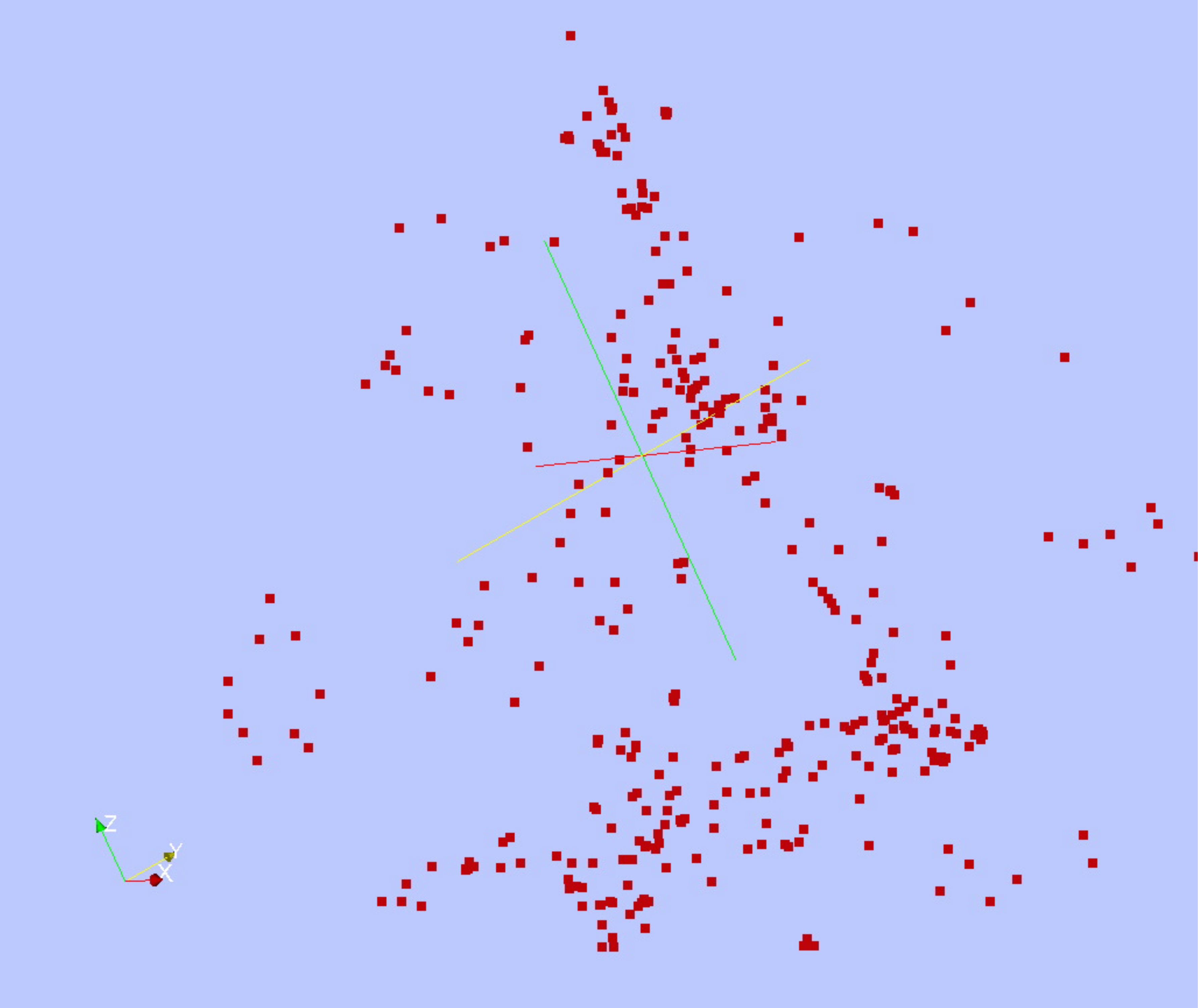}
  \end{center}  
  \caption{A particle representation of the caustics.
    \label{fig:particles}
  }
\end{figure}
\begin{figure}
  \begin{center}
    \includegraphics[scale=0.3]{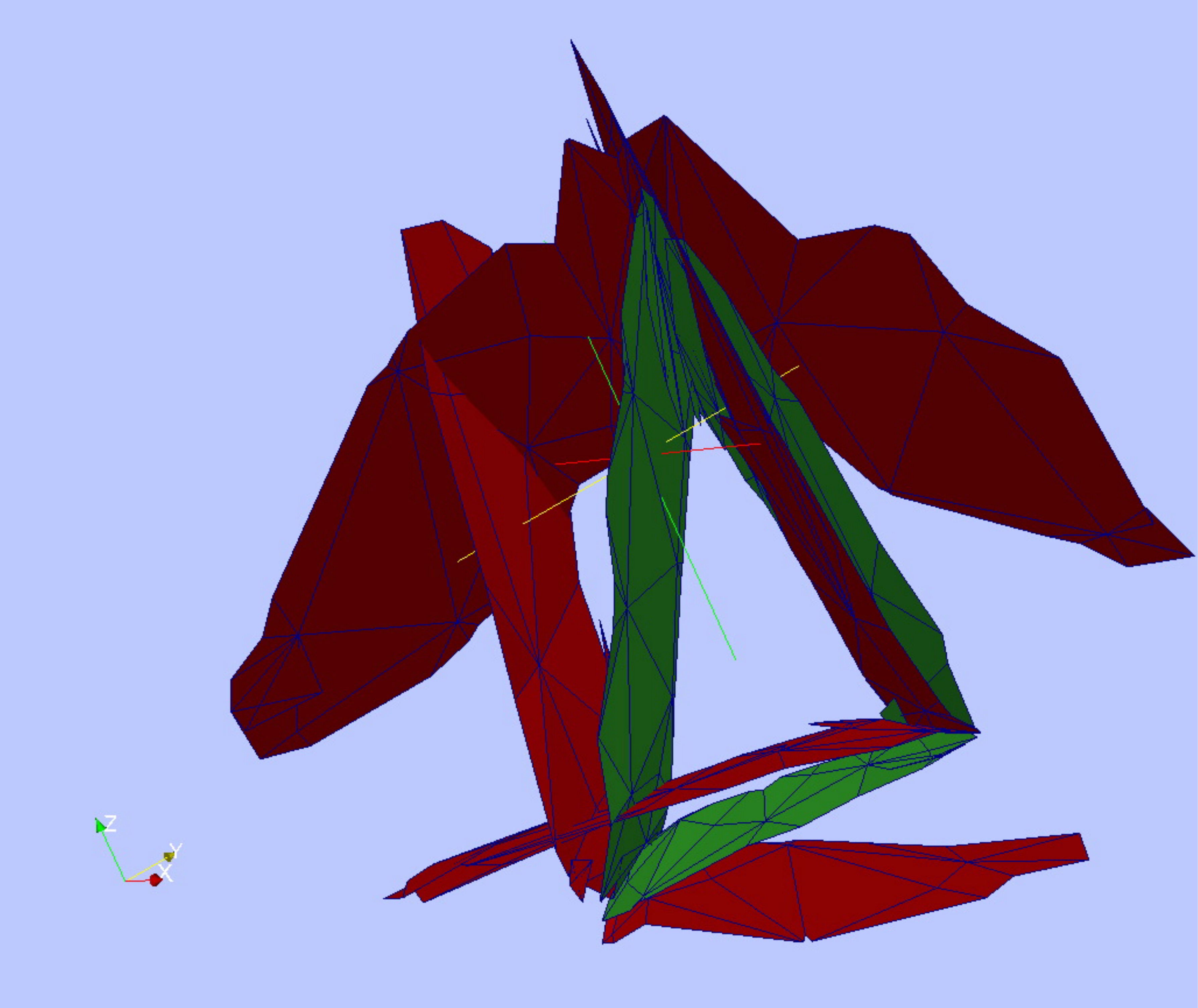}
  \end{center}  
  \caption{Tessellation representation of the same structure as shown by the particles.
    \label{fig:tessellation}
  }
\end{figure}

Probing the caustic structure in gravitationally bound halos is a
difficult problem which has not been properly addressed until rather
recently.  White \& Vogelsberger \cite{whi09} proposed a method to detect
caustic crossings in a running simulation, watching the deformation of
small mass elements around particles at each timestep.  Very recently,
Shandarin et al.\ [2012] and Abel et al.\ [2011] more explicitly
tracked the dynamics of the phase-space sheet, crucially using a tessellation within it.  This allows greater progress to be made from a single snapshot.
Although the methods differ in some details, the basic idea was the
same in both studies.  The advantage of a tessellation technique
over a particle representation is illustrated in Figures
\ref{fig:particles} and \ref{fig:tessellation}. Figure
\ref{fig:particles} shows the particles lying on two caustic surfaces
related to $\lambda_1$ and $\lambda_2$ .  Figure
\ref{fig:tessellation} using exactly the same information as the dot
plot shows two families of caustics in red and green.  Now the
particles are treated as the vertices of the tessellation of the phase
space sheet.  In the next section we briefly explain the method.

\section{Tessellations within the dark-matter sheet for studying caustics}
\label{sec:lalgorithms}
In this section, we explore tessellations that are useful for finding
and analyzing two-dimensional creases in three-dimensional space.

First, though, we briefly review some techniques that work purely in
Eulerian space to detect structures.  These have the clear advantage
that they can be applied to observations in principle, in which the
initial conditions are not known.  Some of these Eulerian techniques
are entirely local, depending on the local arrangement of mass
\cite{HahnEtal2006,Miguel2007,SousbieEtal2008,Sousbie2011}.  Another
approach is global, defining voids to tessellate space.  For instance,
voids can be defined as density depressions outlined by a watershed
transform \citep{Platen2007,Neyrinck2008}.  In this framework, walls,
filaments, and haloes are defined according to where voids meet each
other, and the dimensionality of borders separating them
\citep{Miguel2010}.  Another dynamical algorithm for void-finding
\cite{LavauxWandelt2010} estimates orbits of particles in the final
conditions, designed to detect structures in the primordial density
field.

The first step in the Lagrangian-tessellation algorithm for studying
caustics is the `triangulation' of Lagrangian space itself. The
uniform cubic mesh often used for generating initial positions and
velocities for the N-body simulations is triangulated by subdividing
each cubic voxel of the mesh into five tetrahedra.  The vertices of
these tetrahedra are the particles being tracked through the
simulation, which can be alternatively thought of as vertices of a
mesh covering the phase-space sheet.  The tetrahedra represent the
fluid elements that continuously fill the space. The mass particles
moving in the course of the gravitational evolution deform the
tetrahedra but do not fracture the continuity of the three-dimensional
phase-space sheet. The tetrahedra change their parity every time they
experience collapse in a two-dimensional triangle. Keeping the initial
order of the vertices in each tetrahedron, one can identify the change
of parity by the change of the sign in the volume of the tetrahedron
as computed with a determinant.
 
The next step is to select the triangle faces shared by two
neighboring tetrahedra with opposite parities. This completes the
triangulation of the caustic surfaces at every time step.
Figs.\ \ref{fig:pancakes} and \ref{fig:structure} show the evolution
of the structure described by the Zel'dovich approximation.  In this
first-order theory, by Eq.\ (\ref{eq:volume}), the parity of a fluid
element can be changed at most three times.  Given initial Gaussianity
(which holds to a good approximation), one can easily find the
statistics of parity evolution that is determined by the probability
density function of three eigenvalues, $\lambda_1, \, \lambda_2, \,
\lambda_3$. For instance, no more than about 92\% of all fluid
particles may experience one parity transition since in about 8\% of
the initial volume $\lambda_1$ is negative \cite{dor70}.  In the real
world as well as in cosmological N-body simulations the number of
parity changes is enormous.

Although the caustics are the boundaries between the regions with
different number of streams there is no direct general relation
between the number of streams and number of parity transitions due to
nonlocal character of structure evolution described by mapping. For
example, the interior part of the red cusp in Fig.\
\ref{fig:eig12} above the green caustic contains a different number of
streams than the interior part lying below the green caustic. The
particles lying on the crossing line of two caustics came from very
different parts of Lagrangian space and their paths could be extremely
weakly related.

These new numerical techniques allow a deeper insight into
the complex nonlinear evolution of the large scale structure
in the universe. An example of a useful outcome of the studies of
caustics is a unique definition of physical voids as the regions of
one-stream flows. An N-body simulation of the 'standard' $\Lambda$CDM
model showed that the total volume occupied by physical voids is about
93\% of the total volume \cite{ShandarinEtal2012}.

Such tetrahedra also allow density estimates within the dark-matter
mesh \citep{AbelEtal2011}.  The densities within each stream can then
be added up to a density estimator in many ways more robust than
Eulerian density estimates that ignore the initial arrangement of
particles.  The most common density estimates in cosmology are
Eulerian, volume-weighted density estimates, for example counting the
number of particles in each cell of a cubic grid.  There are also
mass-weighted density estimates, returning the density separately at
each particle, for example using a Voronoi or Delaunay tessellation \cite{svdw,voboz,vdws}.  These are usefully parameter-free and
adaptive, but a truly Lagrangian density estimate \citep{AbelEtal2011}
has the potentioal to be even more physically meaningful, defining the
density at each position as a sum over Lagrangian streams.  This would
have minimal dependence on particle discreteness, and could prove
quite useful to implement within $N$-body simulations.  Although
particle discreteness seems not to be a major issue usually, in some
circumstances (such as warm-dark-matter simulations, where small-scale
clustering is suppressed), artificial structures appear from particle
discreteness \citep{WangWhite2007}.  In such cases, this truly Lagrangian
density estimate could be essential for accuracy.

Tracking the parity as in these Lagrangian-tessellation methods allows
collapsed structures to be detected, but does not immediately give
their morphology (i.e.\ whether they are pancakes, filaments or
haloes).  Keeping track of the axes along which particles cross each
other gives this extra information \cite{FalckEtal2012}, in an
algorithm called \org\footnote{{\centerline{Order-ReversIng}
    \centerline{Gravity, Apprehended} \centerline{Mangling Indices}}}.
In the one-dimensional halo of Fig.\ \ref{fig:spirals}, a natural
place to put the boundary of the structure is at the transition
between where one and three streams overlap when projected to the $x$
axis.  In three dimensions, structures are classified according to how
many perpendicular axes particles within them have been crossed along
by other particles.  Particles in voids, walls, filaments and haloes
have been crossed along 0, 1, 2, and 3 perpendicular directions.  This
is a conveniently parameter-free, objective, geometrical, and
dynamical identification of structures and placement of their
boundaries.  However, this simple particle-crossing criterion does not
distinguish substructures from larger structures.  Such subhaloes,
coils in six-dimensional phase space, likely require a more
sophisticated algorithm to detect.

\section{Conclusion}
A natural tessellation of the final, observed matter and galaxy
distribution (known as Eulerian space) is into voids, as explored
elsewhere in this volume.  Tessellations in Eulerian space are also of
great use for measuring quantities such as densities and velocities
adaptively.  In this contribution, we described tessellations of the
initial dark-matter sheet (known as Lagrangian space).  The natural
tessellation in this case is like an origami crease pattern, dividing
the sheet into streams bordered by folds, or caustics.  Also as in the
Eulerian case, tessellations are useful for analysis, for detecting
stream-crossings and estimating densities natively within the
dark-matter sheet.

\section*{Acknowledgments}

The authors thank Miguel Arag\'{o}n-Calvo for the use of the
simulations presented in Figs.\ \ref{fig:morphparity1} and
\ref{fig:morphparity0}, and for many discussions; Robert Lang for an
inspiring colloquium about paper origami, and valuable discussions;
Tom Hull for permission to use Fig.\ \ref{fig:twocolor}; Eric Gjerde
for permission to use Fig.\ \ref{fig:vorigami}; and Bridget Falck,
Alex Szalay, Rien van de Weygaert and Oliver Hahn for discussions.
MCN is grateful for support from the Gordon and Betty Moore
Foundation.

\bibliographystyle{springer}
\bibliography{refs}

\end{document}